# Surface acoustic wave lasing in a silicon optomechanical cavity

J. Zhang[1,*], P. Nuño-Ruano[1], X. Le Roux[1], M. Montesinos-Ballester[1], D. Marris-Morini[1], E. Cassan[1], L. Vivien[1], N. D. Lanzillotti-Kimura[1], and C. Alonso-Ramos[1,*]

[1]Université Paris-Saclay, CNRS, Centre de Nanosciences et de Nanotechnologies, 10 Boulevard Thomas Gobert, 91120 Palaiseau, France

* Corresponding authors: jianhao.zhang@c2n.upsaclay.fr
carlos.ramos@c2n.upsaclay.fr

**Integrated optomechanical cavities stand as a promising means to interface mechanical motion and guided optical modes. State-of-the-art demonstrations rely on optical and mechanical modes tightly confined of in micron-scale areas to achieve strong optomechanical coupling. However, the need for tight optomechanical confinement and the general use of suspended devices hinders interaction with external devices, limiting the potential for the implementation of complex circuits. Here, we propose and demonstrate a new approach for optomechanical cavities coupling free-propagating surface acoustic waves (SAWs) and guided optical modes. The cavity is formed by a periodic array of silicon nanopillars with subwavelength separation, implemented in silicon-on-insulator substrate. Optical pumping yields a strong radiation pressure that drives the harmonic vibration of the pillars, periodically deforming the silica under-cladding and exciting the SAW. The propagation of the SAW deforms the cavity period, modulating the resonance wavelength to close the optomechanical coupling loop. Based on this concept, we experimentally demonstrate a phonon laser at room temperature and ambient conditions with optical pump power as low as 1 mW. We also show the possibility to cascade this process, achieving a frequency comb generation with more than 30 harmonic lines. These results open a new path to achieve strong bidirectional coupling between integrated waveguides and SAW, with a great potential for a wide range of applications in quantum and classical domains.**

## Introduction

Integrated optomechanical cavities allow precise control of optical and mechanical modes, yielding tight confinement in micron-scale areas and strong photon-phonon interactions [1]. Indeed, the ability to modulate the intensity of the optical signal by coupling to mechanical modes, make optomechanical cavities a promising solution for the implementation of microwave-photonic oscillators. Two approaches have been considered: i) photon lasing, when photon lifetime exceeds phonon lifetime, and ii) phonon lasing, when phonon lifetime exceeds photon lifetime. Optomechanical photon lasing, also referred as Brillouin lasing, yields Stoke signals with ultra-narrow linewidth [2]. For instance, bulk whispering gallery mode resonators exploited cascading of this process to generate 21.7 GHz microwave signal with phase noise of only -110 dBc at 100 kHz offset [3]. More recently, Brillouin photon laser has been recently demonstrated in suspended silicon resonators [4]. In the approach based on phonon lasing, the phonon noise is substantially reduced, and can even reach the Schawlow–Townes limit, when the Stokes signal is a frequency shifted copy of the pump [5]. Hence, beating of different Stoke signals in a photodetector can yield low-phase noise microwave signal generation. Phonon lasing has already been achieved in optomechanical cavities based on different materials, including silicon [6], gallium arsenide [7] and lithium niobate [8]. All these demonstrations required suspended structures. As an example, silicon suspended optomechanical crystal cavities yielded cascading of the phonon lasing process to generate a frequency comb with frequency spacing of 3.9 GHz [6]. Despite these encouraging demonstrations, the need for tight confinement of optical and mechanical modes in micron-scale suspended cavities and the use of suspended structures hampers their interconnection with other photonic or mechanical devices, compromising the potential for their integration in complex circuits.

Conventional silicon-on-insulator (SOI) photonic waveguides, as well as waveguides made with emerging photonic technologies like silicon nitride (SiN), thin film lithium niobate on insulator (LNOI) and aluminum nitride on insulator (AlNOI) yield no phonon confinement due to the high stiffness of the waveguide material, compared to the silica cladding. An initial experimental attempt to yield phonon confinement in SOI waveguides has been reported [9], based on phase-velocity reduction by narrowing the Si waveguide core. Non-suspended optomechanical devices has been demonstrated using guiding materials with lower stiffness than the cladding, like chalcogenide integrated waveguides [10] and lithium niobate on sapphire [11]. Yet, the comparatively lower optomechanical interaction strength, precluded the demonstration of optomechanical photon or phonon lasing in non-suspended planar circuits.

Here, we propose a new approach for the implementation of non-suspended optomechanical cavities achieving strong interaction between guided optical modes and free-propagating surface acoustic waves (SAWs). Based on this concept, we implement an optomechanical resonator in SOI technology, without suspended structures, experimentally demonstrating phonon lasing at room temperature and ambient conditions by optical excitation and readout using integrated silicon waveguides with a pump power of 1 mW. SAWs can be routed over millimeter distances with extremely low dissipation, while allowing efficient electromechanical coupling through interdigital transducers (IDTs). In addition, the use standard SOI technology, i.e. of non-suspended integrated waveguides, substantially facilitates interconnection with key optoelectronic circuits, including integrated modulators and photodetectors. Hence, our approach opens a new route for the implementation of complex circuits combining electromechanical, optomechanical and optoelectronic devices, monolithically integrated on a silicon chip.

## Results

**Optomechanical cavity combining guided optical modes and surface acoustic waves**

Figure 1a shows the schematic view of the proposed SOI optomechanical cavity, comprising a periodic array of silicon segments with a constant period ($\Lambda$) and gap length ($L_g$). The width of the pillars is parabolically reduced and increased to form an optical cavity. The cavity is optimized to yield resonant optical modes with quasi-transverse-electric (TE) polarization, along the $x$ axis, with wavelength near 1550 nm. The optical modes exhibit a strong longitudinal component ($E_z$ in Fig. 1b) that is tightly confined within the air gaps, exerting a strong radiation pressure on the silicon segments. The optical radiation pressure produces a harmonic motion of the stiff silicon segments, periodically deforming the soft silica under-cladding, thereby exciting the SAWs. SAW propagation along the cavity modulates the resonance wavelength, closing the optomechanical excitation loop.

The optimized cavity has a silicon thickness of $t$ = 500 nm, a period of $\Lambda$ = 300 nm and a gap length of $Lg$ = 100 nm. The buried-oxide layer has a thickness of 3 µm. These dimensions are fully compatible with standard silicon technology available in CMOS foundries. The center of the cavity, the silicon segments have a width of Wc = 800 nm, while the segment width at the cavity sides is Ws = 1000 nm. We consider input and output strip waveguides with a width of $W_{WG}$ = 1000 nm and a 100-µm-thick silicon substrate.

The full photonic structure, comprising cavity and access waveguides, supports a wide number of SAW modes, with frequency ranging from a few megahertz to a few tens of gigahetrz. As an example, Fig. 1c shows the calculated displacement profile of SAWs with frequencies of 34 MHz, 127 MHz and 347 MHz. The SAW deforms the cavity, locally stretching and compressing the period of the cavity, leading to a modulation of the resonance frequency.

The calculated optomechanical coupling rate, $g_o$, between the first-order optical mode and SAWs modes with different frequencies is presented in Fig. 1d. A high coupling rate of up to 80 kHz is achieved for a SAW mode with frequency of 24 MHz. These results show that the coupling rate decreases for increasing SAW frequency.

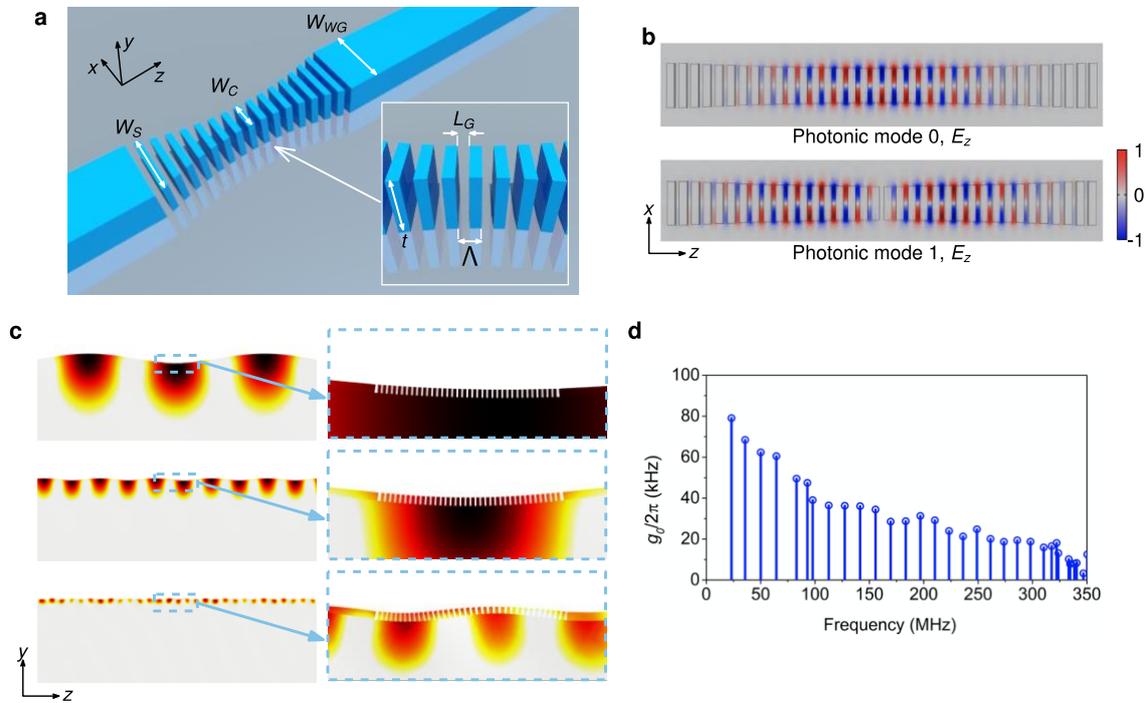

**Figure 1 | Calculated performance of the proposed SOI optomechanical cavity.** **a** Schematic view of the resonator geometry. The period and gap length are constant while the width is parabolically reduced towards the cavity center. Both ends of the cavity are connected to access strip waveguides. **b** Calculated mode electric field distribution of the first two optical modes of the cavity. **d** Calculated displacement profile of SAW modes with frequencies of 34 MHz, 127 MHz and 347 MHz. **e** Optomechanical coupling rate between to the first-order optical mode and SAW modes with different frequencies.

## Phonon laser and frequency comb source

The dynamical optomechanical backaction is experimentally characterized using the setup presented in Fig. 2a. Optical transmission and reflection are measured using an optical spectrum analyzer (OSA). The RF response of the reflected signal is recorded using a photodetector connected to an electronic spectrum analyzer (ESA). Measurements are performed at room temperature and atmospheric pressure.

The device is fabricated using classical silicon photonics process, including electron-beam lithography and reactive ion etching. All dimensions of the structure are compatible with Si technology available in CMOS foundries. Figure 2b shows scanning microscope images of the fabricated cavity. Fiber-chip grating couplers are used to inject optical excitation and to measure the optomechanical cavity response. The first-order optical mode, with a wavelength of 1553 nm, has a measured quality factor of $1.553 \times 10^4$. This device is optically pumped using a tunable laser coupled to an erbium-doped fiber amplifier, yielding 1 mW power coupled into the access silicon waveguide. To probe the mechanical modes, we fix the optical power and scan the wavelength near the optical resonance from a blue-detuned position. Note that the intra-cavity power increases as the pump wavelength approaches the cavity resonance.

Figure 2c shows the evolution of the RF spectrum of the signal backscattered from the optomechanical cavity, recorded as a function of the intracavity power. We maintain blue-detuned driving conditions for all intracavity-powers considered here. For intra-cavity power above 30 mW, a series of high-order harmonics are obesrved in the RF spectrum. These harmonics result from a cascading effect where Stokes and anti-Stokes photons, generated by optomechanical coupling, serve as new pump signals to create additional Stokes and anti-Stokes photons [6]. This process is analog to Brillouin cascading in waveguides [12]. Up to 30 harmonic lines are observed, expanding over a 900 MHz bandwidth.

Figures 2d and 2e show the measured RF power and linewidth as a function of the intracavity power, for the first spectral harmonic line with a frequency near 50 MHz. Phonon lasing threshold is achieved for an intracavity power of 12.4 mW (corresponding to a 1 mW of optical power coupled to the access waveguide) with a 45 dB increase in the RF power and linewidth reduction of 4 orders of magnitude from tens of MHz to the kHz range. Once within the phonon laser regime, the mechanical mode has a linewidth of $\Gamma_m/2\Pi$ = 8 kHz. Figure 2f compares the spectral response of the first harmonic below and above the lasing threshold, showing a frequency shift from 50 MHz to 35 MHz.

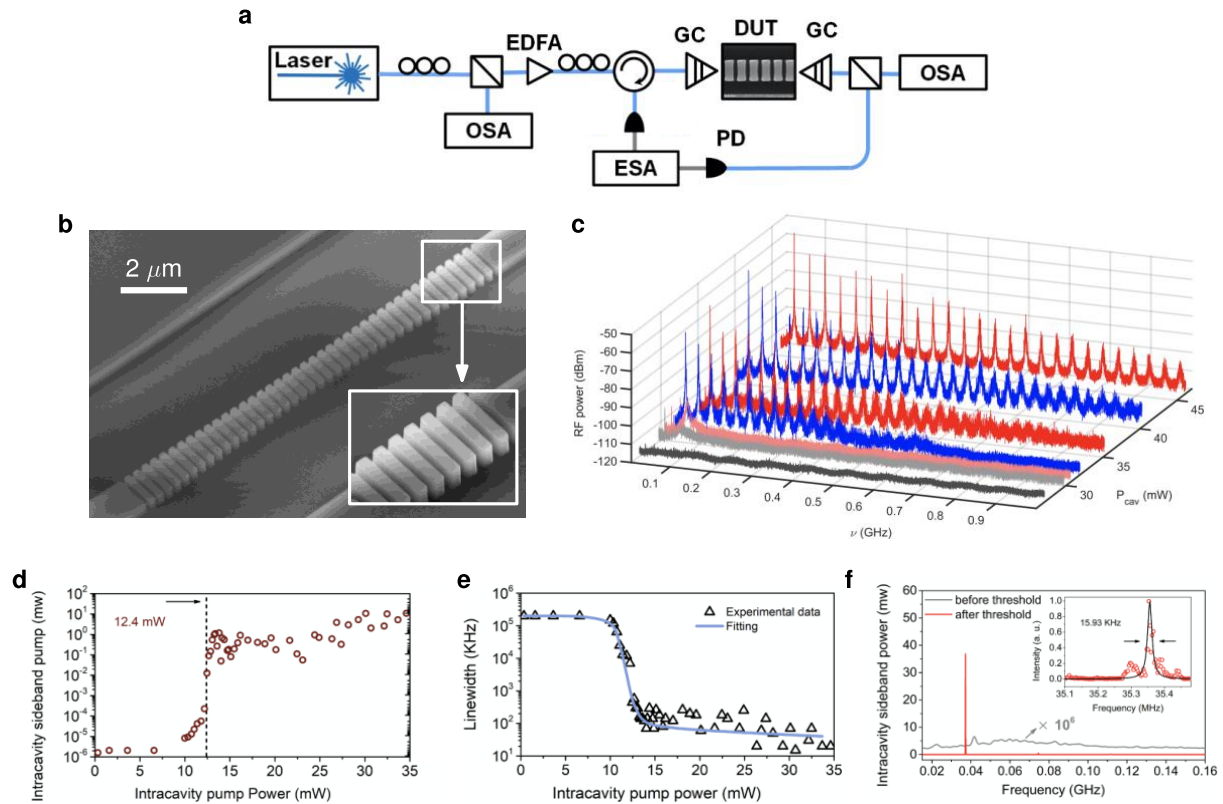

**Figure 2 | Lasing and frequency comb. a** Experimental setup used for optomechanical characterization. ESA: electronic spectrum analyzer. OSA: optical spectrum analyzer. GC: grating coupler. DUT: device under test. PD: photodetector. **b** Scanning microscope image of the fabricated optomechanical cavity. **c** Evolution of the radio-frequency spectrum, as a function of the intracavity power ($P_{Cav}$). Spectrum is recorded by blue-detuned pumping with optical power in the access waveguide fixed at 1 mW. Evolution of the first harmonic RF line, with frequency near 50 MHz as a function of the intra-cavity power: **d** RF power and **e** linewidth. **f** Comparison of the spectral shape of the first harmonic line below and a above the lasing threshold.

## Conclusions

We have proposed and demonstrated a new type of optomechanical cavity bridging guided optical modes and SAWs. We have experimentally demonstrated, phonon lasing with an optical pump power as low as 1 mW and a signal to noise ratio of 60 dB, illustrating the strong optomechanical coupling of the proposed approach. In addition, we have shown the generation of up to 30 harmonic lines by cascading of the optomechanical coupling. The cavity is implemented on silicon-on-insulator technology without the need for suspended structures, facilitating future integration with integrated optoelectronic devices like modulators and photodetectors. On the other hand, SAWs can be routed over millimeter distances with negligible dissipation, opening interesting prospects to interface our cavities with piezoelectric transducers. This demonstration opens a new path for the future development of opto-electro-mechanical devices, with applications in optical metrology, microwave frequency synthesizing, and microwave quantum transduction.

## Methods

### Optomechanical simulations

Optomechanical modeling is performed using the COMSOL Multiphysics solution. The considered SOI wafer comprises a 500-nm-thick top silicon layer, a 3-µm-thick buried oxide layer and 100-µm silicon substrate. We use Bloch-Floquet boundary conditions.

### Device fabrication and experimental characterization

For the fabrication of the SOI microresonators we used electron beam lithography (Nanobeam NB-4 system, 80 kV) with ZEP520A photoresist and dry etching with an inductively coupled plasma etching (SF6/C4F8) for structure definition.

## Acknowledgements
The fabrication of the device was performed at the Plateforme de Micro-Nano-Technologie/C2N, which is partially funded by the Conseil Général de l'Essonne. This work was partly supported by the French RENATECH network.